# Electro-optically programmable photonic circuits enabled by wafer-scale integration on thin-film lithium niobate

Yong Zheng[1,2], Haozong Zhong[1,2], Haisu Zhang[2], Rongbo Wu[2,7], Jian Liu[1,2], Youting Liang[1,2], Lvbin Song[1,2], Zhaoxiang Liu[2], Jinming Chen[2], Junxia Zhou[1,2], Zhiwei Fang[2], Min Wang[2], and Ya Cheng[1,2,3,4,5,6,*]

[1]*State Key Laboratory of Precision Spectroscopy, East China Normal University, Shanghai 200062, China*

[2]*The Extreme Optoelectromechanics Laboratory (XXL), School of Physics and Electronic Science, East China Normal University, Shanghai 200241, China*

[3]*State Key Laboratory of High Field Laser Physics and CAS Center for Excellence in Ultra-intense Laser Science, Shanghai Institute of Optics and Fine Mechanics (SIOM), Chinese Academy of Sciences (CAS), Shanghai 201800, China*

[4]*Collaborative Innovation Center of Extreme Optics, Shanxi University, Taiyuan 030006, China.*

[5]*Collaborative Innovation Center of Light Manipulations and Applications, Shandong Normal University, Jinan 250358, China*

[6]*Shanghai Research Center for Quantum Sciences, Shanghai 201315, China*

[7]rbwu@phy.ecnu.edu.cn

*ya.cheng@siom.ac.cn

**Abstract:** Programmable photonic circuits performing universal linear-optical transformations underpin vital functions in photonic quantum information processing, quantum-enhanced sensor networks, machine learning and many other intriguing applications. Recent advances in photonic integrated circuits facilitate monolithic integration of externally controlled Mach-Zehnder interferometers which can implement arbitrary unitary transformation on a large number of input/output modes. In this work, we demonstrate a 4×4 programmable linear photonic circuit on lithium niobate on insulator platform employing fast, power-efficient and low-loss electro-optical phase shifters, showing enormous advantages in terms of configuration rate and power consumption. Our device is capable of fast switching with 500 ps rise time and 1.7 ns fall time, and possesses a total on-chip power dissipation of only 15 μW when operated at 1 MHz modulation, and an insertion loss of 0.15 dB for each modulator and an on-chip extinction ratio of -34 dB for both cross and bar routes.

## 1. Introduction

Benefiting from their compact footprint, high computation speed, low power consumption, sustainable scalability and high phase stability, programmable linear photonic circuits composed of cascaded Mach–Zehnder interferometers (MZIs) to perform the general special unitary group of degree $N$ (i.e., $SU(N)$) are becoming increasingly important in emerging applications such as photonic quantum information processing, quantum-enhanced sensors, optical neural network and machine learning[1–8]. Stringent requirements are raised by these emerging applications on very large-scale photonic integration (VLSPI) and high-performance on-chip components, such as optical modulators with extremely low losses, ultra-high switching speed, and high power-efficiency. Owing to the maturing complementary-metal–oxide–semiconductor (CMOS)-compatible very large-scale integration (VLSI) processes of high scalability and low manufacturing cost, material platforms including silicon-on-insulator (SOI), silicon nitride, and indium phosphide (InP) have been prevalently investigated with proof-of-concept demonstration [1–3,9–13]. Despite the great progresses made so far on such highly scalable material platforms, unprecedented challenges in maintaining a reasonable power consumption and propagation loss become more and more difficult to meet with the ever-increasing scale of programmable photonic circuits. The relatively high power consumption combined with the relatively slow tuning speed roots in the lack of intrinsic electro-optic effect in most semiconductor material platforms, making it necessary to use thermo-optic or piezo-optomechanical tuning for realizing the phase modulation in the fabricated photonic circuits [12].

Due to its excellent electro-optic and nonlinear optical properties, relatively high refractive index and wide transparency window, lithium niobate ($LiNbO_3$, LN) has been pursued as an ideal material for photonic applications since the 1960s. Formidable challenges exist for decades in obtaining high-quality LN thin films as well as forming high-quality photonic structures during LN dry etching process with the quest for dense and high-performance photonic integration. With the advent of commercially available thin-film LN wafers prepared by ion slicing and the rapid advances in thin-film LN fabrication techniques, an abundant of optical components have been developed on the lithium niobate on isolator (LNOI) platform[14–19], including high-performance meter-scale long waveguides with propagation losses as low as 3 dB/m[20,21] and high-speed efficient electro-optical modulators with tens of gigahertz bandwidth and low power consumption below 1 fJ bit[-1] [16]. As compared to the established CMOS-compatible material platforms, LNOI fulfills all the stringent device requirements raised by programmable photonic circuits by its nature. Wafer-scale LNOI photonic integrated circuits conforming to the requirement on VLSPI have been fabricated with deep ultraviolet lithography (DUV) while maintaining low optical propagation loss of 27 dB/m [22]. Meanwhile, photolithography assisted chemo-mechanical etching (PLACE) technology [23], which patterns the hard mask with maskless femtosecond laser direct writing and then etches the LN with chemo-mechanical polishing (CMP), has also emerged as a promising technique for VLSPI LNOI devices production. PLACE is capable of manufacturing wafer-scale device with high fabrication uniformity, competitive production rate and very low optical propagation loss (i.e. ~2.5 dB/m), enabled by the synergetic actions of high average power femtosecond laser and large-range motion stage together with the extremely smooth interface etched by the CMP process [21].

In this work, we demonstrate a 4×4 programmable linear photonic circuit on LNOI platform fabricated using the PLACE technology. Our device is capable of achieving fast switching of 500 ps rise time and 1.7 ns fall time, total on-chip power dissipation of only 15 μW when operated at 1 MHz modulation, an insertion loss of 0.15 dB for each MZI-unit and an on-chip extinction ratio below -34 dB for both cross and bar routes using Miller's calibration algorithm [24]. The demonstrated LNOI based programmable photonic circuit outperforms its counterparts based on other material platforms in essential functional features such as response time, insertion loss and power consumption, holding great promise in emerging classical and quantum applications.

## 2. Design principle and device realization

Fig. 1(a) schematically illustrates the programmable linear photonic circuit composed of an array of MZI units (1) to (6). Using a two-dimensional subspace within a four-dimensional Hilbert space $[D_{MZI}]_{4\times4}$ to define the unitary transformation matrix of the MZIs (1) to (6), a SU(4) can be constructed. The device illustrated in Fig. 1(a) constructs a transformation matrix $[D]_{4\times4}$ that linearly converts the input signal vector $[I]_{4\times1}$ to the output signal vector $[O]_{4\times1}$. Fig. 1(b) exemplifies the experimental implementation of one MZI-unit, featuring a tandem connection of two MZIs with an intermediate phase shifter ($\theta$) and an output phase shifter ($\varphi$). The two MZIs (BSL and BSR) play the role of adjustable directional couplers controlled by internal phase shifters ($\theta_L$ and $\theta_R$). Following Miller's algorithm, such tandem MZI design allows delicate tuning of the non-perfect splitting of directional couplers subject to fabrication errors to the perfect 50-50 splitting ratios, thereby a perfect *SU*(2) transformation is realized by each MZI-unit despite the existence of a global transmission loss.

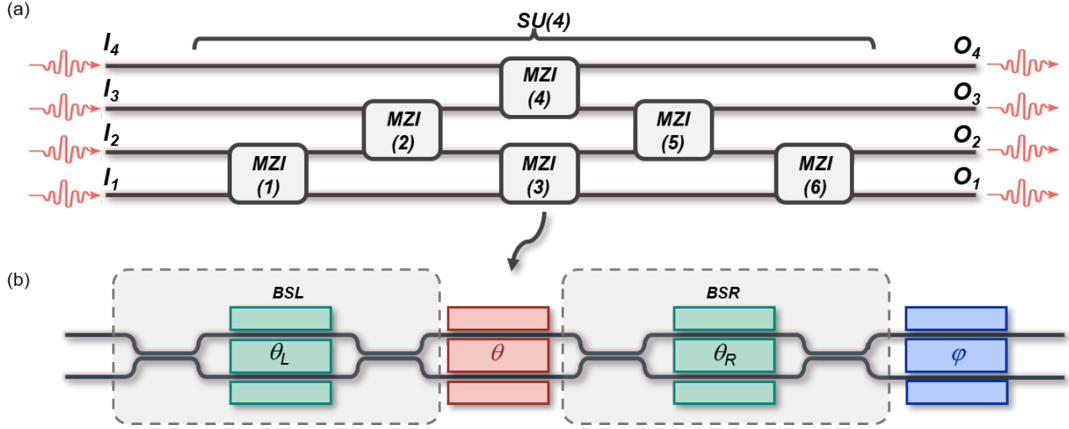

Fig. 1. (a) The arbitrary SU(4) composed of MZI units (1) to (6) that construct the 4×4 programmable linear photonic circuit. (b) Example MZI unit which functions as a perfect SU(2) using Miller's algorithm.

Our device is fabricated on a X-cut 4-inch LNOI wafer with a 500-nm-thick LN layer bonded to a buried silica (SiO$_2$) layer on the 500-μm-thick silicon support (NANOLN) using PLACE technology. The details of the fabrication process can be found in our previous work [21,25]. The photograph of the fabricated programmable linear photonic circuit captured with a digital camera is shown in the center of Fig. 2(a). The arbitrary *SU*(*4*) transformation is realized by 6 reconfigurable MZI-units including 24 phase modulators which can be addressed individually. In our device, all phase shifters utilize electro-optical (EO) actuation of LN with crystal Z-axis lying in the horizontal plane and being oriented perpendicular to the waveguides. With such a configuration, the ground-signal-ground (GSG) electrodes will exert an electric field with equal strength but opposite direction in the two arms of each MZI when applying a voltage, resulting in a phase difference between light transmitted in the two arms.

The internal structure of each MZI-unit is further illustrated in Fig. 2(b), where each phase modulator is approximately 7 mm long and the total length of the MZI-unit is ~30 mm. It has been measured in our experiment that the optical loss of a single directional coupler is ~0.02 dB, and the propagation loss of the optical waveguides is ~0.025 dB/cm, giving rise to the total optical loss of 0.15 dB for a reconfigurable MZI-unit. For confining the *SU*(*4*) transformation circuits within the 4-inch wafer, the device is folded four times over a 32 mm × 8 mm chip area. The relatively large device footprint compared to SOI platforms is mainly due to the relatively low voltage efficiency of the purely linear EO modulator based on Pockels effects and the moderate LN-to-silica refractive index contrast. This is a trade-off for the overwhelming advantages of electro-optical effects in terms of configuration rate and power consumption. It should be pointed that there is no sacrifice in transmission loss despite our MZI is nearly two orders of magnitude longer compared to those on the best performing devices based SOI [2], thanks to the low optical absorption of LN and the extremely smooth waveguide sidewalls produced by the PLACE technology. On the other hand, by taking the advantage of PLACE technology's unique ability on the meter-scale continuous lithography, ultra-large-scale chip processing with extremely low optical loss is now possible.

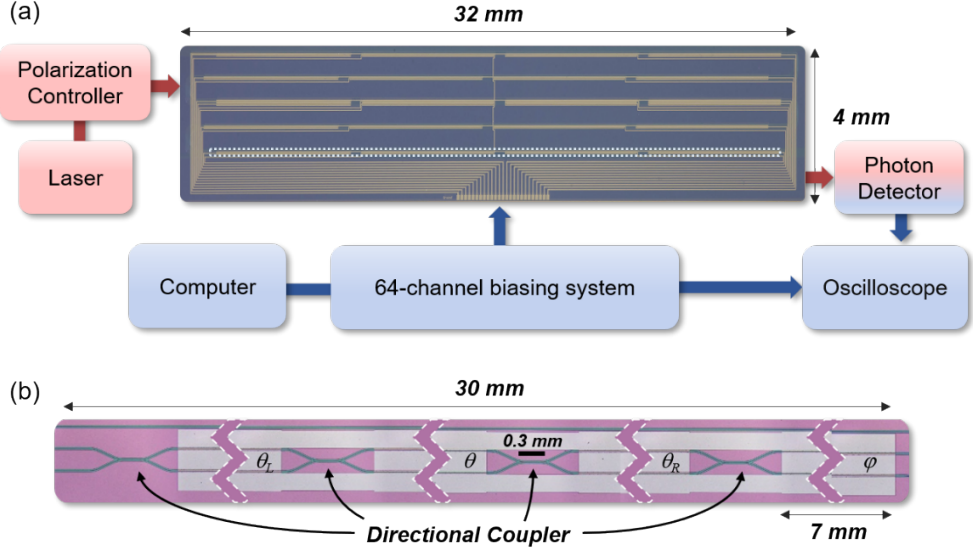

Fig. 2. (a) The arbitrary SU(4) transformation consisting of 24 phase modulators that can be tuned individually and 6 reconfigurable MZI-units. (b) The microscope picture of an entire programable MZI unit with extra phase shifters $\theta_L$ and $\theta_R$.

## 3. Device calibration and characterization

The fabricated programmable photonic circuit normally features random phase offsets and non-perfect splitting ratios due to the inevitable fabrication process fluctuations. Thus, all the MZIs in the device must be calibrated before being programmed for a particular application. The calibration process is carried out based on the topology and order of the MZI units in the device with the experimental setup illustrated in Fig. 2(a). Details about the experimental setup is described in Methods. As shown in Fig. 3(a), the whole process for calibrating the internal phase shifters begins with MZI-units (4), (5), and (6) (red box), and then MZI-units (2) and (3) (blue box) before MZI-unit (1) (green box). The detail of the calibration process can be found in Methods as well. Table 1 presents the required bias voltages for the cross state and bar state configurations of the MZI-units in the 4×4 programmable linear photonic circuit.

Table 1 Required bias voltages for the cross state and bar state configurations of MZI units in the 4x4 programmable linear photonic circuit.

| MZI unit | | (1) | (2) | (3) | (4) | (5) | (6) |
|---|---|---|---|---|---|---|---|
| $\theta$ | $V_{B.S.}$ (V) | -8.08 | -1.03 | -3.97 | -2.51 | 1.87 | -5.68 |
|  | $V_{C.S.}$ (V) | -2.38 | 4.00 | 2.81 | 3.12 | 6.51 | -1.09 |
| $\varphi$ | $V_{B.S.}$ (V) | -2.46 | -0.52 | -5.66 | 4.03 | -1.28 | - |
|  | $V_{C.S.}$ (V) | 4.78 | 4.46 | -0.52 | 8.99 | 3.21 | - |

To characterize the reconfiguration speed of the fabricated device, we selected the path $I_4$-$O_3$ and the MZI unit (5) and measured the transmission versus voltage plot as shown in Fig. 3(b). The result shows a half-wave voltage ($V_\pi$) of 4.64 V, as well as extinction ratios of -36 dB and -34 dB for the bar route ($I_4$-$O_3$) and the cross route ($I_4$-$O_2$), respectively. When the MZI-unit (4) was set in the cross state, a 100 MHz square-wave voltage with a peek-to-peek voltage of 4.4 V generated by an arbitrary waveform generator (AWG) and amplified by an electronic amplifier was exerted on the internal phase shifter of MZI-unit (5). Fig. 3(c) shows the measured curve of transmission versus time, indicating a rise time of 500 ps and a fall time of 1.7 ns as evidenced in Fig. 3(d) and (e), respectively. The degraded performance in high-frequency response compared to similar devices based on LNOI is mainly due to the impedance mismatch and the large microwave loss of the employed electrodes, which are directly fabricated by femtosecond laser micromachining in order to maintain the high fabrication efficiency and large footprint. The fabricated electrodes are of a thickness of only 100 nm and not specifically designed for high-frequency operation, and the microwave signal needs to travel a long distance of 20 mm in the electric wire until reaching the modulators. It has been well established that with carefully designed travelling wave electrodes, a 3-dB bandwidth above 100 GHz can be readily achieved in LNOI devices [16,18,19], providing enough room to promote the computation power by raising the tuning speed of the MZIs in the future.

The power consumption of EO phase shifters can be analyzed by considering charging and discharging of a capacitor consisting of the GSG electrodes and optical waveguides. The energy dissipated per charge/discharge process can be estimated as $\Delta E_{DISS} = CV_{DD}^2/2$ [26], in which $C$ is the capacitance of the phase shifters and $V_{DD}$ is the charging voltage. In our device, the capacitance $C$ is calculated as 21 fF, and the $V_{DD}$ is the same as the average half-wave voltage ($V_\pi$) of the phase shifters which is 5.38 V according to Table 1, resulting in a power dissipation of 304 fJ for each π-phase shift. So the power consumption for a single phase shifter can be estimated as 0.6 μW when operating at 1 MHz modulation rate, and the power consumption for the matrix transformation device which contains 24 phase shifters in total can be estimated to be 15 μW. It is noteworthy that by carefully designing the GSG electrodes and the waveguides, the power consumption of each phase shifter can be further lowered to ~0.37 fJ bit$^{-1}$ when operating at 70 Gbit s$^{-1}$ [16].

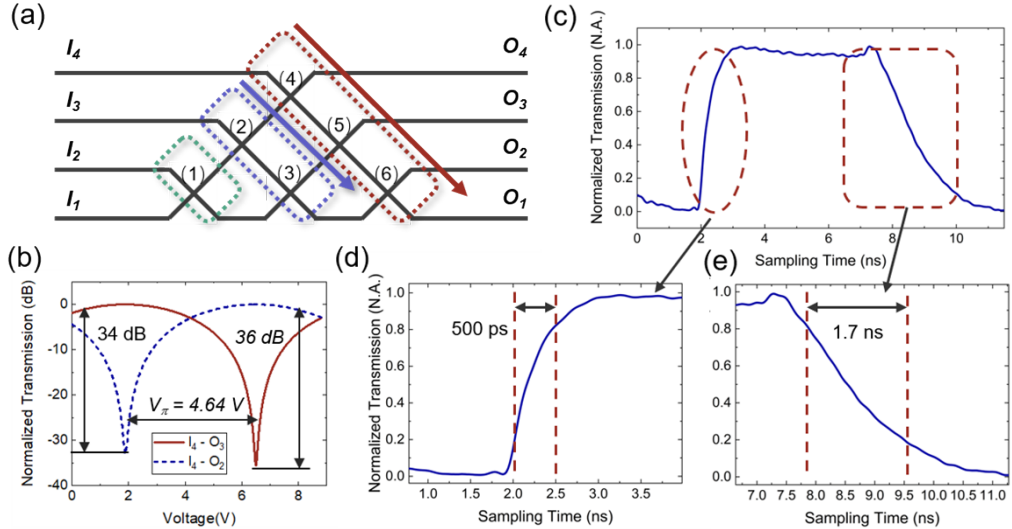

Fig. 3. (a) Schematic of the characterization order of the MZI-units in the *SU(4)*. (b) Normalized optical power of the 2×2 reconfigurable MZI unit (5) from I$_4$ input ports to O$_3$ and O$_2$ output ports as function of bias voltages. (c) Normalized transmission of route I$_4$-O$_3$ with a 100 MHz, 4.4 V peak-to-peak (pk–pk) square wave applied on MZI unit (5). (d) and (e) Zoom-in view of the rising and the falling edges of the transmission curve, highlighting the rise time of 500 ps and the fall time of 1.7 ns.

## 4. Demonstration of the linear matrix transformation and discussion

The programming process of a linear optical processor based on the 4×4 reconfigurable MZI as an arbitrary *SU(4)* is performed by decomposing the linear transformation matrix $[T_{SU(4)}]$ into the unitary matrices of the corresponding MZI-units in the device. The decomposition process is accomplished by successively multiplying $[T_{SU(4)}]$ with $[D_{MZI}^{(i)}]^{-1}_{H_{4\times 4}}$ in a specific order beginning from the right-handed side, in which $[D_{MZI}^{(i)}]^{-1}_{H_{4\times 4}}$ is the inverse of the unitary transformation matrix of a specific MZI unit. As illustrated in Fig. 4(a), the decomposition process starts from MZI-units (1), (2), and (4) (see, the red box in Fig. 4(a)) to null all off-diagonal elements in the fourth row and the fourth column of the decomposed matrix by which the effective dimension can be reduced by one. After that, multiplication of the transmission matrix of MZI-units (3), (5) (the blue box in Fig. 4(a)) and finally (6) (the green box in Fig. 4(a)) will null all the off-diagonal elements, which results in an identity matrix. The details of the decomposition process can be found in Methods. As an example, consider programming two different SU(4)s with two random linear transformation matrixes:

$$[T_{SU(4)_1}] = \begin{pmatrix} -0.076 - 0.82j & -0.059 - 0.361j & -0.097 - 0.214j & -0.251 + 0.264j \\ -0.249 - 0.105j & 0.665 + 0.209j & -0.209 - 0.584j & -0.055 + 0.231j \\ -0.356 + 0.313j & -0.142 - 0.514j & 0.274 - 0.299j & -0.500 - 0.277j \\ 0.041 - 0.15j & -0.249 + 0.183j & -0.518 - 0.358j & 0.073 - 0.692j \end{pmatrix} \#(1)$$

$$[T_{SU(4)_2}] = \begin{pmatrix} 0.085 + 0.05j & 0.214 - 0.681j & -0.552 - 0.418j & -0.034 + 0.016j \\ -0.546 - 0.529j & 0.405 + 0.166j & -0.369 + 0.241j & 0.152 + 0.118j \\ -0.011 - 0.537j & 0.019 - 0.387j & 0.37 + 0.111j & -0.515 - 0.383j \\ -0.352 - 0.026j & -0.346 + 0.171j & -0.135 - 0.399j & 0.235 - 0.703j \end{pmatrix} \#(2)$$

The required phase shifts for the phase shifters of the MZI-units in the device to implement the two SU(4)s were obtained by the decomposition process mentioned above. Table 2 lists the required phase shifts and the corresponding bias voltages for each MZI-units.

Table 2 Calculated phase shifts and the corresponding bias voltages of the MZIs for programing the SU(4)s

| MZI | $SU(4)_1$ | | | | | | $SU(4)_2$ | | | | | |
|---|---|---|---|---|---|---|---|---|---|---|---|---|
| | (1) | (2) | (3) | (4) | (5) | (6) | (1) | (2) | (3) | (4) | (5) | (6) |
| $\theta$ (Rad) | -2.68 | 2.08 | -0.34 | 1.76 | -1.85 | -0.65 | 1.73 | 1.16 | -1.36 | -1.69 | -2.67 | 1.63 |
| $V_\theta$ (V) | 4.16 | 7.32 | 2.08 | 6.27 | 3.78 | 7.15 | 0.75 | 5.85 | 13.43 | 0.10 | 4.82 | 1.29 |
| $\varphi$ (Rad) | 0.92 | -0.21 | -0.79 | 0.17 | 0.61 | -0.53 | 0.11 | 0.97 | -0.27 | -0.55 | -1.74 | -0.09 |
| $V_\varphi$ (V) | 6.91 | 4.13 | 8.46 | 9.26 | 4.09 | - | 5.04 | 6.00 | 9.32 | 8.12 | 0.73 | - |

The performance of our device was further examined by setting the bias voltages according to the parameters listed in Table 2 and measuring the optical power at each output port when the incident light enters from the input ports (1)-(4) one by one. The normalized optical power from all the different path can be expressed as a 4×4 transmission matrix, where each element of the matrix should be equal to the modulus of the corresponding element of the transmission matrix of the target SU(4). The theoretical and experimental results were both plotted in Figs. 4(b)-4(e). Fig. 4(f) shows the difference between the theoretical and experimental results for all the optical paths. The maximum deviation of the experimental value from the theoretical prediction is ~8% as can be clearly seen from Fig. 4(f). The highest contribution to the experimental deviations comes from the fluctuation induced by the vibration of the lensed fibers during the calibration and testing process, which can be further optimized with mode convertors integrated on the input and output ends of the programmable photonic circuit.

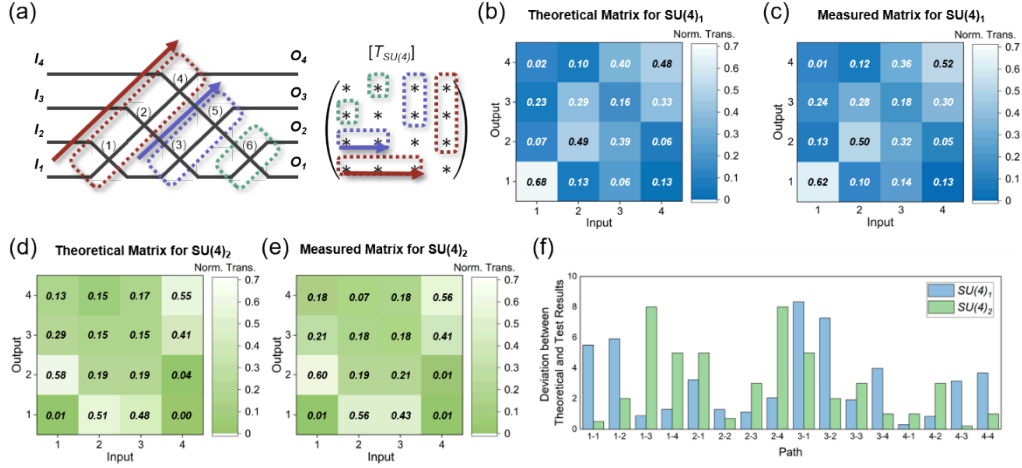

Fig. 4. (a) Schematic illustration of the MZI-units order in the SU(4) section for decomposing the unitary transformation matrix $[T_{SU(4)}]$. (b) and (c) Theoretical and measured transmission matrix of the normalized optical power for all possible paths in the device to construct $SU(4)_1$. (d) and (e) Theoretical and measured transmission matrix of the normalized optical power for all possible paths in the device to construct $SU(4)_2$. (f) Difference between the theoretical and experimental results for all possible paths.

Table 3 presents a comparison of the performance of our device to those of the existing state-of-the-art devices based on SOI and silicon nitride platforms. It can be clearly seen that the LNOI programable integrated circuit outperforms the existing devices built upon the SOI and silicon nitride platforms in terms of optical loss, extinction ratio, power consumption and reconfiguration time. In addition, it is noteworthy that our device is able to achieve extinction ratios exceeding 34 dB using Miller's calibration algorithm, even with the relatively large fabrication errors given by the large footprint of the device. The high tolerance of fabrication imperfections is particularly advantageous for practical applications associated with the VLSPI, where high-precision fabrication may not always be feasible. Last but not least, the scalability of our approach appears optimistic as the performance of the device can be reliably promoted by increasing the number of MZIs to construct larger optical network on the chip and incorporating travelling-wave electrodes to achieve multi-hundred GHz operation bandwidth. Both approaches will take the device to an unprecedented performance level for both quantum and classical optical computation.

**Table 3 Comparison of key performance metrics for programable integrated circuits on different material platforms.**

|  | SOI | | | SiN | LNOI |
| --- | --- | --- | --- | --- | --- |
|  | [2] | [10] | [11] | [12] | This work |
| Modes Number | 26 | 8 | 4 | 4 | 4 |
| Phase Shifter Number | 176 | 56 | 20 | 12 | 24 |
| Footprint | 4.9 × 2.4 mm² | - | - | 1.3 × 0.6 mm² | 32×4 mm² |
| MZI Loss | 0.1 dB | 0.4 dB | 1.5 dB | 3.5 dB | 0.15 dB |
| Extinction Ratio | -60 dB | -27 dB | -43 dB @ B.S. <br> -15 dB @ C.S. | -30 dB | -34 dB @ C.S <br> -36 dB @ B.S. |
| π-Shift Power Consumption | 55 mW | 3 mW | 21 mW | 200 μW @ 1MHz | 0.6 μW @ 1MHz |
| Switch Time | 10 μs | 50 μs | 30 μs | 5 ns | 500 ps Rise <br> 1.7 ns Fall |

## 5. Conclusion

To conclude, we have demonstrated the fabrication and characterization of a 4×4 programmable linear photonic circuit on the LNOI platform using the PLACE technology. The device exhibits excellent performance with fast switching of a rise time of 500 ps and a fall time of 1.7 ns, an insertion loss of 0.15 dB for each MZI unit and an on-chip extinction ratio below -34 dB for both cross and bar routes. The total on-chip power dissipation is only 15 μW when operated at 1 MHz modulation. Our results demonstrate that LNOI programmable integrated circuits demonstrate superior performance compared to SOI and silicon nitride devices in terms of optical loss, extinction ratio, power consumption, and reconfiguration time. This work provides a significant step towards the realization of large-scale, low-power-consumption, and high-performance photonic circuits for cutting-edge classical and quantum applications.

## 6. Methods

*Forming an arbitrary SU(N) with cascaded MZIs.*

The unitary transformation matrix of the MZI units associated with the input and output signals through the respect waveguide channel *m* and *n* within the *SU(N)* can be expressed as:

$$[D_{MZI}(m,n)]_{H_{N \times N}} = \begin{pmatrix} 1 & 0 & \cdots & \cdots & \cdots & \cdots & 0 \\ 0 & 1 & \cdots & \cdots & \cdots & \cdots & 0 \\ \vdots & \vdots & \ddots & \cdots & \vdots & \vdots & \vdots \\ 0 & 0 & \cdots & je^{j\frac{\theta}{2}}e^{j\phi}sin\left(\frac{\theta}{2}\right) & je^{j\frac{\theta}{2}}e^{j\phi}cos\left(\frac{\theta}{2}\right) & \cdots & 0 \\ 0 & 0 & \cdots & cos\left(\frac{\theta}{2}\right) & -sin\left(\frac{\theta}{2}\right) & \cdots & 0 \\ \vdots & \vdots & \cdots & \cdots & \cdots & \ddots & \vdots \\ 0 & 0 & \cdots & 0 & 0 & \cdots & 1 \end{pmatrix} \begin{matrix} 1 \\ 2 \\ \vdots \\ m \\ n \\ \vdots \\ N \end{matrix} . \#(1)$$

The unitary transformation matrix $[D]_{4 \times 4}$ can be obtained by the product of the unitary transformation matrices $[D]_{MZI}$ of MZIs (1) to (6) of which the mathematic form is determined by the order of MZIs in the photonic circuit. Therefore, the unitary transformation matrix $[D]_{4 \times 4}$ is defined as follow:

$$[D]_{4 \times 4} = [D_{MZI}^{(6)}]_{H_{4 \times 4}} \cdot [D_{MZI}^{(5)}]_{H_{4 \times 4}} \cdot [D_{MZI}^{(4)}]_{H_{4 \times 4}} \cdot [D_{MZI}^{(3)}]_{H_{4 \times 4}} \cdot [D_{MZI}^{(2)}]_{H_{4 \times 4}} \cdot [D_{MZI}^{(1)}]_{H_{4 \times 4}} . \#(2)$$

*Details of the experimental setup*

A polarization controller (Polarization Synthesizer, PSY-201, General Photonics Corp., Chino, California, USA) was used to rotate the light field direction at a wavelength of 1550 nm provided by a continuous laser (CTL 1550, TOPTICA Photonics Inc., Farmington, New York, USA) to transverse-electric (TE) polarization, since the device is designed to operate only with TE-polarized light for the best EO-modulation efficiency. Light is coupled in and out of the fabricated device via lensed tapers, and is finally converted to electronic signals using photon detector (APD-2M-A-100K, Luster, Beijing, China), which can be displayed and analyzed on a high-speed real-time oscilloscope (MSO64B, Tektronix Inc., Beaverton, USA). The device is electronically programmed using a 64-channel biasing system controlled by an analog output module (PXIe-6739, National Instruments Corp., Austin, Texas, USA).

*Details of the calibration process*

The whole process for calibrating the internal phase shifters begins with MZI units (4), (5), and (6) (red box), and then MZI units (2) and (3) (blue box) before MZI unit (1) (green box), as shown in Fig. 3(a). As an example, we first calibrate MZI unit (4) by applying bias voltages to its internal phase shifter $\theta_4$, as well as the corresponding extra phase shifters $\theta_{4L}$ and $\theta_{4R}$. Optical transmission powers through the paths $I_4$-$O_4$ and $I_4$-$O_3$ were recorded versus the bias voltage applied on the internal phase shifter. The extinction ratios of the paths $I_4$-$O_4$ (bar route) and $I_4$-$O_3$ (cross route) can be both optimized by repeatedly adjusting the voltages applied on the extra phase shifters $\theta_{4L}$ and $\theta_{4R}$, and we can define the bar-state voltage $V_{B.S.}$ by the applied voltage at which the transmission through the bar path is maximum. Likewise, and the cross-state voltage $V_{C.S.}$ can be defined by the applied voltage at which the transmission through the cross path is maximum. Afterwards, the bias voltage required for an internal phase shifter to produce a desired phase shift $\theta$ can be calculate as:

$$V(\theta) = V_{C.S.} + \frac{\theta |V_{B.S.} - V_{C.S.}|}{\pi} \#(3)$$

To calibrate the outer phase shifters, the corresponding MZI unit and the subsequent MZI unit will be set as 50:50 beam splitters to form another MZI, and the MZI units on each arm of the newly formed MZI will operate in the bar state. For example, during the calibration process of the phase shifter $\varphi_2$, MZI units (2) and (5) will be set as 50:50 beam splitters and MZI units (4) and (3) will operate in the bar state. Phase shifter $\varphi_4$ can also be calibrate using the same protocol considering the configuration of the push-pull electrodes. Using the same process, $\varphi_3$ and $\varphi_5$ can be calibrated in the MZI formed with MZI units (3) and (6) by operating MZI unit (5) in the bar state. Despite the fact that $\varphi_6$ cannot be calibrated in our device, this does not affect the final test results, since it only introduces an unknown constant phase to the output light.

*Details of the decomposition process*

It is essential to point out that we can always find a $[D_{MZI}^{(i)}]^{-1}_{H_{4\times 4}}$ such that in each step of the successive multiplications of $[T_{SU(4)}]$ with $[D_{MZI}^{(i)}]^{-1}_{H_{4\times 4}}$, an off-diagonal element in the lower triangle of the resultant matrix becomes zero, i.e., the multiplication of an arbitrary $[T_{SU(4)}]$ with $[D_{MZI}^{(1)}]^{-1}_{H_{4\times 4}}$ can be expressed as follows:

$$[T_{SU(4)}] \cdot [D_{MZI}^{(1)}]^{-1}_{H_{4\times 4}} = \begin{pmatrix} U_{11} & U_{12} & U_{13} & U_{14} \\ U_{21} & U_{22} & U_{23} & U_{24} \\ U_{31} & U_{32} & U_{33} & U_{34} \\ U_{41} & U_{42} & U_{43} & U_{44} \end{pmatrix} \#$$

$$\cdot -je^{-j\frac{\theta_1}{2}} \begin{pmatrix} e^{-j\varphi_1}\sin(\frac{\theta_1}{2}) & \cos(\frac{\theta_1}{2}) & 0 & 0 \\ e^{-j\varphi_1}\cos(\frac{\theta_1}{2}) & -\sin(\frac{\theta_1}{2}) & 0 & 0 \\ 0 & 0 & 1 & 0 \\ 0 & 0 & 0 & 1 \end{pmatrix} \#(4)$$

$$= \begin{pmatrix} * & * & * & * \\ * & * & * & * \\ * & * & * & * \\ 0 & * & * & * \end{pmatrix}, \quad if: \theta_1 = 2\arctan\left(\frac{-U_{42}}{U_{41}}\right).$$

As illustrated in Fig. 4(a), the decomposition process starts from MZIs (1), (2), and (4) (red box in Fig. 4(a)), which will null the first, second and third element in the fourth row of the decomposed matrix, respectively. Because of the unitary property of the decomposed matrix, when all off-diagonal elements become zero, the off diagonal elements in the corresponding column also become zero in each row. Thus, the effective dimension of the decomposed matrix can be reduced by one as follow:

$$[T_{SU(4)}] \cdot [D_{MZI}^{(1)}]^{-1}_{H_{4\times 4}} \cdot [D_{MZI}^{(2)}]^{-1}_{H_{4\times 4}} \cdot [D_{MZI}^{(4)}]^{-1}_{H_{4\times 4}} = \begin{pmatrix} & & & 0 \\ ( & T_{SU(3)} & ) & 0 \\ & & & 0 \\ 0 & 0 & 0 & 1 \end{pmatrix} \#(5)$$

The next step is to multiply the transmission matrix of MZIs (3) and then (5) to reduce the effective dimension of the resultant matrix to two dimensions (blue box in Fig. 4(a)). Finally, the multiplication of the decomposed matrix by $[D_{MZI}^{(6)}]^{-1}_{H_{4\times 4}}$ will null all the off-diagonal elements, resulting in an identity matrix (green box in Fig. 4(a)). Thus,

$$[T_{SU(4)}] \cdot [D_{MZI}^{(1)}]^{-1}_{H_{4\times 4}} \cdot [D_{MZI}^{(2)}]^{-1}_{H_{4\times 4}} \#$$
$$\cdot [D_{MZI}^{(4)}]^{-1}_{H_{4\times 4}} \cdot [D_{MZI}^{(3)}]^{-1}_{H_{4\times 4}} \cdot [D_{MZI}^{(5)}]^{-1}_{H_{4\times 4}} \cdot [D_{MZI}^{(6)}]^{-1}_{H_{4\times 4}} = [I]_{4\times 4}\#(6)$$

Equation (5) is equivalent to equation (1), therefore, all the required phase shifts to implement any arbitrary *SU*(*4*) are obtained after the decomposition process.

**Funding.** National Key R&D Program of China (2019YFA0705000), National Natural Science Foundation of China (Grant Nos. 12004116, 12104159, 11874154, 11734009, 11933005, 12134001, 61991444, 12204176), Science and Technology Commission of Shanghai Municipality (NO.21DZ1101500), Shanghai Municipal Science and Technology Major Project (Grant No.2019SHZDZX01), Shanghai Sailing Program (21YF1410400) and Shanghai Pujiang Program (21PJ1403300). Innovation Program for Quantum Science and Technology (2021ZD0301403).

**Disclosures.** The authors declare no conflicts of interest.

**Data availability.** Data underlying the results presented in this paper are not publicly available at this time but may be obtained from the authors upon reasonable request.